\documentstyle[prl,aps,twocolumn,epsfig]{revtex}
\input epsf.sty
\begin{document}
\draft
\flushbottom
\twocolumn[\hsize\textwidth\columnwidth\hsize\csname@twocolumnfalse\endcsname
\title{Theory of Ferromagnetism in Doped Excitonic Condensates}
\author{E. Bascones, A. A. Burkov, and A. H. MacDonald}
\address{Department of Physics, University of Texas at Austin, Austin, Texas 78712}
\date{\today}
\maketitle   
\begin{abstract}

Nesting in a semimetal can lead to 
an excitonic insulator state with spontaneous coherence between conduction
and valence bands and a gap for charged excitations.  In this paper we 
present a theory of the ferromagnetic state that occurs when the density
of electrons in the conduction band and holes in the valence band differ.
We find an unexpectedly rich doping-field phase diagram and an unusual
collective excitation spectrum that includes two gapless collective modes.
We predict regions of doping and external field in which phase-separated
condensates of electrons and holes with parallel spins
and opposing spins coexist.      

\end{abstract}

]

It has long been recognized\cite{Keldysh65,Halperin68}
that a semimetal or small-gap semiconductor could undergo a phase transition
to a state with spontaneous coherence between conduction and valence bands 
and a gap for charged excitations.  When described in terms of 
electrons in the conduction band and holes in the valence band, this state 
can be regarded as an electron-hole pair (exciton) condensate, and its 
mean-field theory 
description is very similar to the BCS mean-field theory of a superconductor.
Excitonic insulator states are expected to occur only if conduction and 
valence bands are adequately nested.

The excitonic insulator state is evidently rare, most likely\cite{vignale} 
because accurate nesting is uncommon and because electron and hole densities are 
usually not equal.  The most compelling example of excitonic-insulator 
behavior to date is that demonstrated by bilayer quantum Hall 
systems\cite{eisensteinrefs}, where a charge gap develops in a state with spontaneous 
coherence\cite{theoryrefs} between electrons in different layers.  
In this paper we consider doped excitonic 
insulators, {\it i.e.} systems in which the conduction-band electron 
density and valence-band hole density differ.  Doping is unfavorable for electron-hole 
pairing, causing the excitonic insulator to evolve first into a metallic excitonic
ferromagnet\cite{Volkov75}, 
and finally into a paramagnetic normal metal state.
In this paper we address properties of the intermediate ferromagnetic state.

This work is motivated in part by the recent
discovery of weak ferromagnetism in lightly-doped divalent
hexaborides \cite{Young99}.
These materials are ferromagnetic in spite of the absence  
of {\it d}- or {\it f}- partially filled orbitals, and have very high Curie 
temperatures 
$T_c > 600 K$ that depend strongly on carrier concentration. 
Early electronic structure calculations had predicted that these materials would be semimetals
or small gap semiconductors \cite{Hasegawa79},
leading Zhitomirsky {\it et al} \cite{Zhitomirsky99} to propose 
excitonic pairing as the source of their ferromagnetism, renewing 
interest in this unusual magnetic 
state\cite{Balents00,Gorkov00,Zhitomirsky00,Balents00b,Ichinomiya01,Gloor00,Murakami01}.  
Uncertainty has been generated, however, by recent experiments\cite{Denlinger} 
which appear to imply that these materials are 
semiconducting with a $\sim 1$eV gap, in agreement with more recent GW approximation  
electronic structure calculations\cite{Bobbert}, casting doubt on the 
excitonic insulator picture.  In this Letter, we present an 
analysis of the collective excitation spectrum, and doping-chemical potential phase diagram
of excitonic insulator ferromagnets, establishing features which could be used to
convincingly identify these states on purely experimental grounds.
In particular, we demonstrate that doped excitonic
condensates are unusual ferromagnets with a characteristic multi-branch 
collective mode structure.

We start by considering a model Hamiltonian ${\cal H}={\cal H}_0+{\cal H}_I$ 
where ${\cal H}_0$ includes both conduction and valence band energies and 
${\cal H}_I$ includes all interactions.
\begin{equation}
\label{2}
{\cal H}_0=\sum_{{\bf k} a \sigma \sigma'}\left[ (\epsilon_{{\bf k} a}-\mu)\delta_{\sigma,\sigma'} - {\bf h}\vec{\tau}_{\sigma,\sigma'} 
\right ]c^{\dag}_{{\bf k} a \sigma}c_{{\bf k} a \sigma'}
\end{equation}
Here ${\bf k}$ denotes the quasimomentum, $\sigma$ the spin, $a=c,v$ is a 
band label, $\mu$ is the chemical potential and ${\bf h}=\frac{1}{2}g\mu_B {\bf H}$ 
here ${\bf H}$ is an external magnetic field.
We start by including only long-range Coulomb interactions, which leads to a model with 
interaction vertices that conserve band labels and spins: 
\begin{equation}
\label{3}
{\cal H}_I=\frac{1}{2\Omega}\sum_{{\bf k} {\bf k'} {\bf q} \sigma \sigma'}
\sum_{a a'}V({\bf q})c^{\dag}_{{\bf k}+{\bf q} a \sigma}
c^{\dag}_{{\bf k}'-{\bf q} a' \sigma'}
c_{{\bf k}' a' \sigma'}c_{{\bf k} a \sigma}
\end{equation} 
where $\Omega$ is the crystal volume.  
This interaction model is known as the dominant term approximation\cite{Halperin68}, 
and is accurate for low-carrier densities and weak Coulomb interactions.   
In this model charge and spin are conserved separately in each band, 
leading to a $SU(2)\times SU(2)$ symmetry \cite{Balents00} that 
will figure prominently in our discussion.  

For simplicity we consider the case of delta-function interactions,
$V({\bf q})=V_0$, and of isotropic, parabolic valence and conduction bands 
that are perfectly nested in the absence of doping:
\begin{equation}
\label{4}
\epsilon_{{\bf k} c}=-\epsilon_{{\bf k} v}=\frac{{\bf k}^2}{2m}-E_G. 
\end{equation}
Here $E_G>0$, the case we consider,  and $E_G<0$ imply semimetallic and
semiconducting behavior, respectively. 

Our results for ground states and excitations are based on unrestricted Hartree-Fock and 
time-dependent Hartree-Fock approximations respectively.
Excitonic condensation is driven by interband exchange terms that 
favor spontaneous phase coherence.  The intraband exchange self-energy 
has a spin-independent contribution which we absorb into 
$E_G$ and an important spin-dependent contribution which has not always been accounted
for in previous work, but plays an essential role in stabilizing
the ferromagnetic state\cite{Balents00}.

Dropping the direct Hartree interaction which plays no role for homogeneous states,
the mean-field Hamiltonian is
\begin{eqnarray}
\label{5}
\nonumber
{\cal H}_{MF}&=&\sum_{{\bf k} a \sigma \sigma'}\left[(\epsilon_{{\bf k} a \sigma}-\mu)
\delta_{\sigma \sigma'}-({\bf h}_a+{\bf h}){\vec \tau}_{\sigma \sigma'}\right] 
c^{\dagger}_{{\bf k} a \sigma}c_{{\bf k} a \sigma'}
\\ 
&-&\sum_{{\bf k} \sigma \sigma'}\left[\left(\Delta_s\delta_{\sigma \sigma'}+
{\vec \Delta}_t{\vec \tau}_{\sigma \sigma'}\right)
c^{\dagger}_{{\bf k} c \sigma}c_{{\bf k} v \sigma'}+h.c.\right] 
\end{eqnarray}
Here 
\begin{equation}
\label{6}
{\bf h}_a=\frac{V_0}{\Omega}{\bf m}_a=\frac{V_0}{2\Omega}\sum_{{\bf k}\sigma\sigma'}
\rho_{aa}^{\sigma \sigma'}({\bf k})\vec \tau_{\sigma'\sigma}
\end{equation}
is the mean-field intraband exchange spin-splitting field (proportional to the band
magnetization ${\bf m}_a$), 
\begin{eqnarray}
\label{7}
\Delta_s&=&\frac{V_0}{2\Omega}\sum_{{\bf k}\sigma\sigma'}
\rho_{cv}^{\sigma \sigma'}({\bf k})\delta_{\sigma\sigma'}\nonumber\\
\vec \Delta_t&=&\frac{V_0}{2\Omega}\sum_{{\bf k}\sigma\sigma'}
\rho_{cv}^{\sigma \sigma'}({\bf k})\vec \tau_{\sigma'\sigma}
\end{eqnarray}
are the singlet and triplet excitonic condensate order parameters,
and
\begin{equation}
\label{8}
\rho_{aa'}^{\sigma\sigma'}({\bf k})=\langle c^{\dag}_{{\bf k}a'\sigma'}
c_{{\bf k}a\sigma}\rangle
\end{equation}
is the Hartree-Fock density matrix.

\begin{figure}
\epsfxsize 9 cm
\centerline{\epsfbox{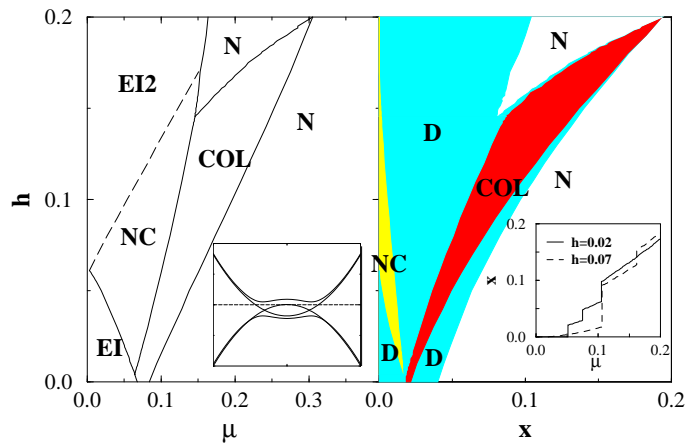}}
\caption{
Phase diagram in the presence of a field as a function of 
chemical potential $\mu$ a) or doping $x$ b) and field $h$ for $E_g=0.1$ and
$V_0=1.1$. Total electronic density is given by $n_0(1+x)$. 
Energy is measured in units of the band width $\pi^2/(2m)=1$. 
The adjacent pure phases occur as phase separated domains in 
the regions labelled D in b).
The inset in a) shows a sketch of the Hartree-Fock quasiparticle bands in the ferromagnetic state.
The inset in b) shows the dependence of doping on chemical potential. 
} 
\label{fig:phasediagram}
\end{figure}

We have solved these equations for a range of chemical potentials and external fields,
allowing for density matrices that have spontaneous coherence between conduction and 
valence bands and either spontaneous or induced spin-polarization in conduction and 
valence bands.  Our results are summarized in Fig.~\ref{fig:phasediagram}.
There are in general several families of solutions, a 
normal state solution (N) in which no symmetries are broken, a ferromagnetic excitonic  
solution in which the excitonic condensate involves only conduction band electrons and 
valence band holes with particular spin orientations (see the inset in Fig. 1a)),
and the excitonic insulator (EI) state in which both spin orientations participate in 
the excitonic condensate.  Invariance of the 
Hamiltonian under independent spin-rotations of conduction and valence band systems 
in the absence of an external magnetic field is key to understanding the phase diagram.
For zero chemical potential and magnetic field a family\cite{Halperin68} of 
symmetry-related EI states is lowest in energy.
For $h=0$, a family of ferromagnetic states forms the ground state over a finite
range of $\mu \ne 0$.  This family
includes states with different spin orientations for condensed conduction
and valence band particles, {\em and} states with different total magnetization magnitude.   
A weak external magnetic field selects from this family the state which has the largest 
magnetization magnitude aligned with the field direction.  
For $\mu \ge 0$ this is always a state in which spin down valence holes are condensed.  
For large $\mu$, where the largest contribution to the conduction band magnetization
comes from uncondensed electrons, it follows that the collinear (COL) state 
which has coherence between parallel spin conduction and valence band electrons, is the ground state.  
For small $\mu$ , on the other hand, the non-collinear state (NC) is selected by a field.
Both COL and NC ferromagnetic excitonic states compete
with the EI state in which all spin orientations are paired, 
and with the N state, with the four different states separated by first order phase transitions.  
Both magnetic and pairing order parameters depend on both magnetic field and chemical potential,
varying continuously within a phase and discontinuously at a phase boundary.
In region EI2 only spin down valence band electrons and spin up conduction band
holes are condensed.  Unlike the NC state, this phase is insulating.

The slopes of the first order phase boundary lines are related to density and 
magnetization discontinuities by the following Clapeyron equation
\begin{equation}
\frac{dh}{d\mu} = - \frac{n_0 \Delta x}{\Delta M}.
\label{eq:clapeyron} 
\end{equation}
Here $x$ is the dimensionless quantity used to measure density  and $n_0$ is 
the density in the absence of doping.
As shown in the inset of Fig.1b) 
$x$ always increases upon crossing these phase boundaries in the 
direction of increasing chemical potential.   
Positive slopes for these phase boundary lines correspond therefore to 
a magnetization that decreases when crossing the phase boundary in the same 
direction.  Inside each phase, density increases monotonically with chemical potential.  
It follows that as a function of density,
the phase diagram consists of pure phase regions interspersed with regions 
of phase separation.  The regime consisting of phase separated COL and NC states, which have 
different magnetizations and excitonic pairing of quite different character, 
is particulary unusual.  Note also the reentrant normal state, predicted earlier\cite{Gorkov00},
induced in the COL by field at large $x$.

Given the unusual nature of these ferromagnets, we can expect that their 
collective mode structure is unusual.  We have evaluated the
dynamic linear response in a time-dependent Hartree-Fock approximation~\cite{colmodeotherstates2}.  
We discuss here only spin-wave collective excitations, those that
 involve tilts of the magnetization orientation  
and are therefore off-diagonal in spin-indices when the valence and 
conduction band mean-field magnetization directions are chosen as quantization axes.  

\begin{figure}
\epsfxsize 9 cm
\centerline{\epsfbox{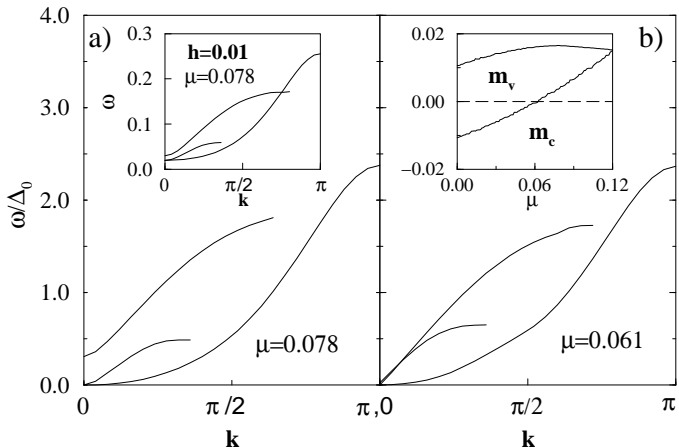}}
\caption{Spin wave energies in units of the $h=0$, $\mu=0$ excitonic insulator order 
parameter $\Delta_0=0.1$ for the COL pure state. 
We measure the momentum in units of the lattice 
parameter. 
These results were 
obtained for $E_g=0.1$ and $V_0=1.1$, in units of the bandwidth. The main figures show zero field results. 
for two different values of the chemical
potential corresponding to positive and zero conduction band magnetizations 
respectively  The inset in b) shows the dependence of the COL state 
valence and conduction band magnetizations on $\mu$. Magnetization is
measured in units of $n_0$. 
The inset in a) shows the collective excitation spectrum at 
finite field.}
\label{fig:colmodes}
\end{figure}

Our results for the spin-wave collective modes $E(q)$ of pure phase 
excitonic  ferromagnets are summarized in
Fig.~\ref{fig:colmodes}. 
For $h=0$, there are {\em two} gapless collective modes with quadratic    
dispersion, corresponding to the independent conduction and valence band broken 
spin rotational symmetries, and an additional soft but gapped collective mode.
An interesting feature occurs in the doping dependence of this 
excitation spectrum.  At an isolated value of $\mu$ 
the spectrum consists of one quadratically dispersing mode and 
two linearly dispersing modes.  
To understand the origin of this odd behavior, it is necessary to 
understand how the separate conduction and valence band contributions to 
the magnetization vary with doping.  
Consider, for example, the collinear state.  
For the electron doping case illustrated here there are no uncondensed valence band holes. 
Uncondensed conduction band electrons, on the other hand, make a contribution to
the magnetization opposite to that of the condensed conduction band electrons,
yielding a net magnetization.
At some point, the overall magnetization from the conduction band vanishes. 
This is point at which the collective excitation spectrum is anomalous, and 
incidentally also the same point, at which COL and NC states have the 
same magnetization.  Apparently the conduction band excitation spectrum at 
this point is much like that of an antiferromagnet, 
with condensed and uncondensed band electrons producing 
opposing magnetization contributions.

\begin{figure}
\epsfxsize 9 cm
\centerline{\epsfbox{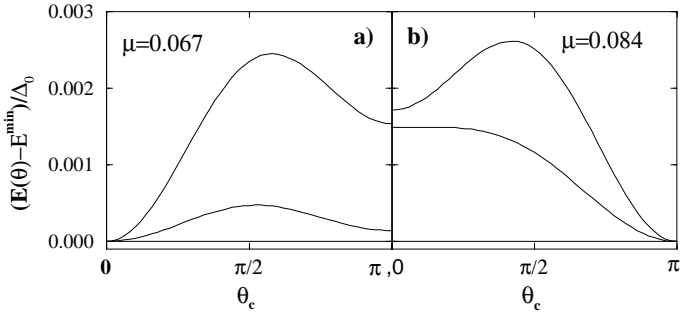}}
\caption{Dependence of the energy on the angle between the spin of the
condensed conduction band and the magnetic field for $\mu=0.067$ and 
$\mu=0.084$ in which the ground state is at finite field in the NC and COL states
respectively, and $\theta_v=\pi$. From bottom to top $h=0.0,0.01,0.02$.}
\label{fig:three}
\end{figure}

In the presence of an external magnetic field our calculations show that 
there are two Larmour 
collective modes at energy $E = 2h=g \mu_B H$.  
At first sight the fact that conduction and valence
collective modes both occur at this energy appears to suggest that their 
moments interact only
with the external magnetic field and not with each other.  
If this were the case, however, each
contribution to the magnetization would simply align with the field and
there would be no energy barrier 
between collinear and non-collinear states.  
Furthermore, it is clear from our microscopic calculations that 
conduction and valence bands are coupled by the combined influences of 
excitonic pairing and Fermi statistics.  
This surprising in-field collective mode behavior can be understood by 
considering the dependence of energy on the orientations of the conduction and 
valence band spins that participate in the pairing.  
In Fig.~\ref{fig:three} we plot, for both 
collinear and non-collinear states, energies obtained from self-consistent 
solutions of 
the mean-field equations for pairing between down spin valence band holes and 
conduction band spins that have polar orientation angle $\theta_c$ with respect to
the magnetic field.  
We see in Fig.~\ref{fig:three} that there is indeed a non-trivial
dependence of the energy on $\theta_c$ with the global minimum occurring at 
$\theta_c=\pi$ in the COL region of the phase diagram and at
$\theta_c=0$ in the NC region of the phase diagram.
Why then do the conduction and valence band collective magnetization 
fluctuations not reflect this dependence on their relative orientations?  
The answer again can be found in the symmetry of Hamiltonian.  
In a magnetic field, the energy of the system is still
invariant under independent rotations of conduction and valence band 
magnetizations around 
the field axis.  
Spin-wave collective excitations are related to the leading second order 
changes in collective energy when magnetization is tilted away from the 
field direction.
Since we are allowed to include only contributions that are invariant 
under separate conduction and valence band spin-rotations, the quadratic coupling term 
$\hat m_{\perp,c} \cdot \hat m_{\perp,v}$, is not allowed.
Higher order terms in the expansion of the magnetization,
for example ones that go like $(\hat m_{\perp,c} \cdot \hat m_{\perp,c}) 
(\hat m_{\perp,v} \cdot \hat m_{\perp,v})$,
and cause interactions between spin-waves, {\em are} symmetry allowed.  
It is interesting to speculate on differences between the role of 
fluctuations in EI ferromagnets, compared to their role in other 
more familiar magnetic states.  At low-temperatures, the Bloch $T^{3/2}$ 
$M(T)$ law should still apply, despite the existence of several low-energy spin-wave branches.
At higher temperatures, excitonic and magnetic order parameters vanish 
simultaneously at the mean-field level.
We expect that mean-field
theory critical temperature estimates should be reliable in the weak-coupling
limit $V_0 \ll E_g$. 

In closing, we remark that in any real excitonic insulator ferromagnet, 
corrections to the dominant term approximation
would partially alter the picture outlined here.
The largest corrections would likely come from interband interaction terms
of the form
$\delta V_0  c^{\dag}_{{\bf k}+{\bf q}c\sigma}c^{\dag}_{{\bf k}'-{\bf q}v
\sigma'}c_{{\bf k}'c\sigma'}c_{{\bf k}v\sigma}$.
This term will produces corrections to the Hartree-Fock energy of the form
$\tilde E^{HF} \propto 2|\Delta_s|^2-{\bf m}_c {\bf m}_v$
We find that the first term is always larger in magnitude,
thus favoring triplet pairing and NC orientation in the ferromagnetic phase.    
The second term favors parallel conduction and valence band magnetizations at 
$h=0$, coupling the two Goldstone modes and gapping one of them.
The resulting $h-\mu$ phase diagram will depend on the competition of magnetic
energy and correction terms.
In the weak coupling limit $\Delta_0 \ll |E_g|$, these terms are expected to 
be negligible and the phase diagram qualitatively unchanged.

We gratefully acknowledge discussions with L. Balents, J. Fern\'andez-Rossier,
Z. Fisk, T.M. Rice and M.E. Zhitomirsky. This work has been supported by 
NSF-DMR0115947 and the Welch Foundation.


\begin{thebibliography}{99}
\bibitem{otherwork2}
Some past work has incorrectly predicted energy differences between states 
that are related by transformations that are in the symmetry group of the
Hamiltonian.
\end{thebibliography}
\end{document}